\documentclass[11pt,a4paper]{article}
\usepackage{amssymb, amsmath, graphicx, subcaption, cite}
\usepackage{float}
\usepackage{graphpap,epsfig,textcomp}
\usepackage{epstopdf}
\usepackage{tabularx}
\usepackage[utf8]{inputenc}
\usepackage[english]{babel}
\usepackage{hyperref}
\usepackage[nottoc]{tocbibind}
\begin{document}

\begin{center}
{\Huge {\bf Phase transitions in the anisotropic 
XY ferromagnet with quenched nonmagnetic impurity}}\end{center}

\vskip 1cm

\begin{center}{\it Olivia Mallick$^1$ and Muktish Acharyya$^{2,*}$}\\
{\it Department of Physics, Presidency University,}\\
{\it 86/1 College Street, Calcutta-700073, INDIA}\\
{$^1$E-mail:olivia.rs@presiuniv.ac.in}\\
{$^2$E-mail:muktish.physics@presiuniv.ac.in}\end{center}

\vskip 1cm

\noindent {\bf Abstract:} The equilibrium behaviours of the anisotropic  XY ferromagnet,
with nonmagnetic impurity, have been investigated in three dimensions by Monte Carlo simulation using
Metropolis algorithm. Two different types of anisotropy, namely,  
the bilinear exchange type and single-site anisotropy are considered here.
The thermodynamic behaviours of
the components of the magnetisations ($M$), susceptibility ($\chi$) and the specific heat ($C$) have been studied 
systematically through extensive Monte Carlo simulations. The ferro-para phase transition has been observed to take place at a lower temperature 
for impure anisotropic XY ferromagnet. The pseudocritical temperature ($T_c^*$) has been found to decrease as
the system gets more and more impure (impurity concentration $p$ increases). In the case of bilinear exchange type of anisotropy
($\lambda$), the pseudocritical temperature ($T_c^*$) increases linearly with $\lambda$ for any given concentration of nonmagnetic impurity ($p$). The slope of this linear
function has been found to depend on the impurity concentration ($p$). The slope decreases linearly with the 
impurity concentration ($p$).
In the case of the single site anisotropy ($D$), the pseudocritical temperature ($T_c^*$) has been found to decrease linearly
with $p$ for fixed $D$. The critical temperature (for a fixed set of parameter values) has been estimated from the 
temperature variation of fourth order Binder cumulants ($U_L$) for different system sizes ($L$). The critical magnetisation
($M(T_c)$) and the maximum value of the susceptibility ($\chi_p$) are calculated for different system
sizes ($L$). The critical exponents for the assumed scaling laws, $M(T_c) \sim L^{-{{\beta} \over {\nu}}}$ and $\chi_p \sim L^{{{\gamma} \over {\nu}}}$, are estimated through the finite size analysis. We have estimated, 
${{\beta} \over {\nu}}$, equals to  $0.48\pm0.05$ and $0.37\pm0.04$ for bilinear exchange and single site anisotropy
respectively. We have also estimated, ${{\gamma} \over {\nu}}$ equals to $1.78\pm0.05$ and $1.81\pm0.05$ for bilinear exchange
and single site anisotropy respectively.

\vskip 3cm


\vskip 2cm

\noindent $^*$ Corresponding author
\newpage

\noindent {\bf {\Large I. Introduction}}

\vskip 0.2cm

\noindent The planar ferromagnet( continuous SO(2) symmetric) in two dimensions, holds a distinctive significance within the realm of phase transitions.
The existence of peculiar kind of ordered phase (vortex-antivortex pair) without any conventional long range ferromagnetic order has made
it special\cite{kosterlitz1, kosterlitz2,betts}. The special kind of phase transitions has been observed in various
thermodynamical systems\cite{jose} namely in two dimensional superconductors, superfluids, liquid-crystals etc..

The planar ferromagnet (classical XY ferromagnet) can give rise to usual phase transition (having long range 
ferromagnetic ordering) in higher dimensions ($d>2$). The systematic investigations of the thermodynamic phase transitions
are done in three dimensional XY ferromagnet\cite{campostrini,hasenbusch}
by Monte Carlo simulations, mainly to estimate the critical temperature and the critical
exponents. The XY universality class has also been specified.

{\it How does the phase transition get affected by anisotropy ?} The XY ferromagnet with single site anisotropy has been
investigated\cite{ma} quantum mechanically (bosonic field) and the critical temperature has been found to increase monotonically with
the strength of single site anisotropy. The XY ferromagnet with bilinear exchange kind of anisotropy has been investigated recently \cite{olivia1}
by Monte Carlo simulation. Here also, the critical temperature has been found to increase linearly with the strength
of bilinear exchange anisotropy. Moreover, the phase transitions in three dimensional XY ferromagnet have been studied\cite{olivia1} with distributed anisotropy. The studies regarding the role of anisotropy are not limited in the thermodynamic phase
transitions only. To study the vortex-loop scaling behaviour, the anisotropic (with different interplane/intraplane coupling ratios) XY ferromagnet
was investigated\cite{biplab} by using the technique of the renormalization group. 
The
renormalization group technique has also been employed\cite{granato} to study the critical behaviours of coupled XY
model.
The Cantor spectra have been observed
\cite{satija} in the one dimensional quasiperiodic anisotropic XY
model. 

The anisotropy may also be introduced by simply varying the type of interactions (ferromagnetic or antiferromagnetic)
in different directions. The quantum critical behaviour of spin-1/2 anisotropic XY model has been investigated\cite{su}
with staggered Dzyaloshinskii-Moriya interaction. The XY model has recently been investigated for various kinds of interactions. Those are higher order
exchange interactions\cite{zuko1}, antinematic kind of interactions\cite{zuko3}, geometrically frustrated interaction\cite{lach1} and higher order
antiferromagnetic interaction\cite{lach2} on a triangular lattice. The phase transition was studied\cite{erol} in the three-dimensional XY
layered antiferromagnet, where the intra-planar interaction is ferromagnetic with inter-planar antiferromagnetic interaction.

In the thermodynamic phase transition, it is quite natural to study the effects of disorder. Apart from the randomly distributed anisotropy\cite{olivia1}, the quenched disorder may also be 
introduced in the system by random fields and randomly quenched nonmagnetic impurity. Recently, the role of
random field in the aging and domain growth has been extensively investigated\cite{puri}. The random fields act as random disorder which usually reduces the critical temperature. On the other hand, the uniform anisotropy generally increases the critical temperature just by breaking the SO(2) symmetry.

A competitive behaviour of the random field and the bilinear exchange anisotropy has recently been investigated\cite{olivia2}
by Monte Carlo simulation. Another kind of random disorder may be 
represented by the randomly quenched nonmagnetic impurity. The site
diluted XY ferromagnet has been studied\cite{filho} by Monte Carlo simulation and found that the phase transition took place
at lower temperatures for higher values of the nonmagnetic impurity. 
This feature is quite natural since the disorder usually reduces the
transition temperature.

However, we have not yet observed any systematic and intensive study on the impure anisotropic XY ferromagnet. {\it It may be a naive question, what will be the critical behaviour of the anisotropic XY ferromagnet in the presence of quenched random nonmagnetic impurity ?} We have addressed this question in this paper.
We have systematically investigated, by Monte Carlo simulation, the effects of nonmagnetic impurity,  both in the bilinear exchange kind of anisotropy as well as in the single site anisotropic XY ferromagnet. 
The manuscript is formatted as follows: the anisotropic XY model is introduced in section-II, the
Monte Carlo methodology is mentioned in section-III, the simulational results are reported in section-IV and the paper
ends with a summary accompanied by some concluding remarks given in section-V.

\vskip 1cm

\noindent {\bf {\Large II. Anisotropic XY ferromagnetic model}}

\vskip 0.2cm

\noindent The  Hamiltonian of anisotropic (bilinear exchange type)  XY ferromagnet is expressed as

\begin{equation}
H= -J \sum_{<i,j>} (1+\lambda)S_i^xS_j^x + (1-\lambda)S_i^yS_j^y
\label{hamiltonian-lambda}
\end{equation}
where $S_{i}^x (=cos \theta_i)$ and $S_{i}^y(=sin \theta_i)$ correspond to  the components of the two dimensional vector (at the i-th lattice site) having unit length, $|S|$=1. $\theta$ is the angle (measured with respect to the positive X-axis)
of the vector $\vec S$. It may be considered as a two dimensional rotor at each lattice site, which can point in any direction (specified by $\theta$).
The $\lambda$ denotes the strength of bilinear exchange anisotropy. For $\lambda=0$, the system becomes conventional isotropic XY ferromagnet and the system represents ferromagnetic XX model for $\lambda=1$ (but $S_i^{x}$ is continuous variable). This anisotropy usually breaks the SO(2) symmetry.
The parameter $J$ is the nearest neighbour (represented by $<ij>$ in the summation) ferromagnetic ($J > 0$) interaction strength. For simlicity, we have considered the uniform ($J$ constant) ferromagnetic interaction, here.
 
One may also think of the anisotropic XY ferromagnet with the
single site anisotropy ($D$) represented by the following Hamiltonian: 

\begin{equation}
H= -J \sum_{<i,j>} S_i^xS_j^x + S_i^yS_j^y -D\sum \left[ (S_i^x)^2-(S_i^y)^2) \right].
\label{hamiltonian-D}
\end{equation}

\noindent Here, $D$ represents the strength of single site anisotropy.

The nonmagnetic impurity is implemented in the anisotropic model by 
considering simply a few site (randomly chosen with probability $p$) having
${\vec S}=0$. The concentration of such nonmagnetic impurity is represented
by $p$. This nonmagnetic impurity ($p$) acts as quenched disorder in the system.

\vskip 1cm

\noindent {\bf {\Large III. Monte Carlo simulation methodology}}

\vskip 0.2cm

\noindent In the present study, a three dimensional simple cubic lattice of size $L$
(=20 here, apart from the finite size analysis)
is considered here. The lattice dimensions of the 
system are three
(simple cubic) and the dimensions of the spin vector are two (planar ferromagnet or XY model). We have applied the conventional periodic boundary 
conditions  in all the three directions of the lattice.

The Monte Carlo simulation starts from a random initial configuration of the spin, corresponding to
a very high temperature phase. This state corresponds to the usual paramagnetic phase (without long range ferromagnetic ordering) having
zero magnetisation.
At a finite temperature $T$ (measured in the unit of $J/k$, where $k$ is Boltzmann
constant), a lattice site (say x,y,z) is chosen randomly 
(at any instant of time $t$) having an random initial spin configuration (represented
by an angle $\theta_i(x,y,z,t)$). A new configuration of the spin 
(at site x,y,z) is also chosen
(represented by $\theta_f(x,y,z,t)$) randomly. The change in energy
($\delta H(t)$) due to the change in configuration (angle)
of spin (from $\theta_i(x,y,z,t)$ to $\theta_f(x,y,z,t)$) is calculated 
from eqn.~(\ref{hamiltonian-lambda}) and eqn.~\ref{hamiltonian-lambda}) for respective cases.  The probability of accepting
the new configuration  is calculated from the Metropolis formula\cite{binder},

\begin{equation}
P_f = {\rm Min}[{\rm exp}({{-\delta H(t)} \over {kT}}), 1].
\end{equation}

A uniformly  distributed (between 0 and 1) random number ($r=[0,1]$) is chosen. 
The chosen site is assigned to the new
spin configuration $\theta_f(x,y,z,t')$ (for the next instant $t'$)
if $r \leq P_f$. In this way, $L^3$ number of sites are updated randomly. $L^3$ number of such
random updates defines a unit time step and is called Monte Carlo step per site (MCSS). The time in this simulation is measured in the unit of MCSS.
Throughout the study the system size $L(=20)$.  The total length of simulation is $2\times10^4$ MCSS, out of which the initial
$10^4$ MCSS times are discarded. Here, to make the system ergodic some initial MCSS are required. All statistical quantities are calculated by averaging over the rest $10^4$ MCSS\cite{binder}. However, for finite size
analysis the length of the simulation is much longer. Each thermodynamic quantities are further averaged over different random realizations 
(typically around 20) of
fixed concentration of impurty ($p$). But for smaller sizes of the systems
($L=10,15$ etc.) the number of such random samples is quite large
(around 200).

\vskip 0.3cm

The instantaneous components of magnetisations are 
$m_{x}(t)= {1 \over {L^3}} \sum s_x (x,y,z,t)\\ 
= {1 \over {L^3}} \sum {\rm cos} (\theta (x,y,z,t))$ 
and 
$m_{y}(t)= {1 \over {L^3}} \sum s_y(x,y,z,t)
= {1 \over {L^3}} \sum {\rm sin}(\theta(x,y,z,t)$. Total instantaneous magnetisation is $m=\sqrt{m_x^2+m_y^2}$.

The equilibrium magnetisation is measured as 
$M= <m>$.
The susceptibility is determined by 
$\chi={L^3\over{kT}}(<m^2>-<m>^2)$. The specific heat has been calculated from $C = {{L^3} \over {kT^2}} (<H^2>-<H>^2)$.
The fourth order Binder cumulant (for system size $L$) has been calculated as $U_L = 1- { {<m^4>} \over {3{<m^2>^2}}}$. The symbol $<..>$, represents the time averaging,
which is approximately (within the length of simulation) equal to the ensemble averaging in the ergodic limit.

\vskip 1.0 cm

\noindent {\bf {\Large IV. Simulational Results}}

\vskip 0.2cm

\noindent In this section, we report the simulational results of our study. The results for the bilinear exchange type of anisotropy ($\lambda$) is given in subsection (a) and that for single site anisotropy ($D$) are reported in subsection (b).

\vskip 0.5cm

\noindent {\bf {\it {(a). Bilinear exchange type of Anisotropy ($\lambda$)}}}

\vskip 0.3cm

\begin{figure}[h]
\begin{center}

\resizebox{10cm}{!}{\includegraphics[angle=0]{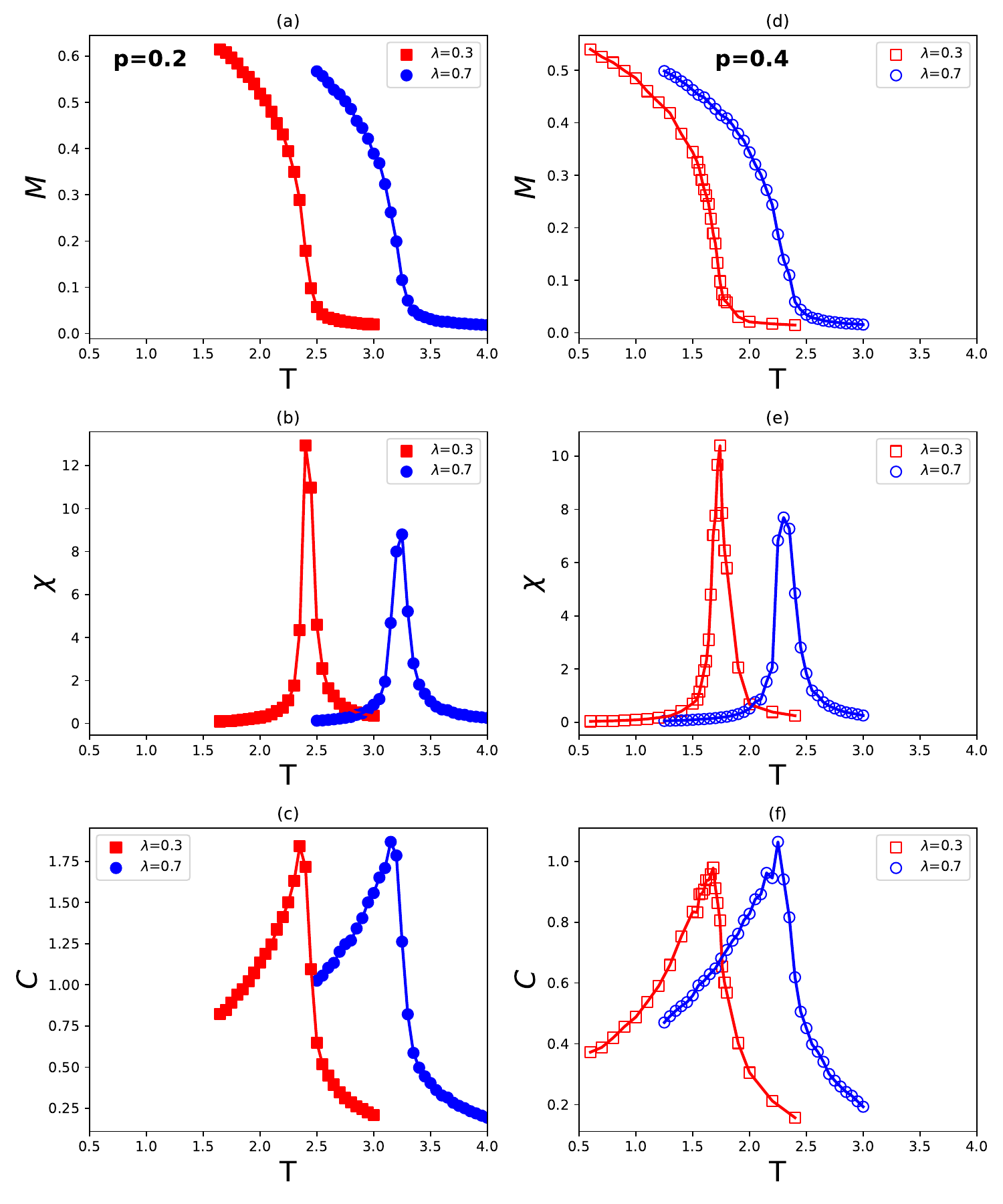}}

\caption{The temperature dependences of all thermodynamic quantities.
In the left panel, for impurity concentration $p=0.2$, 
(a) Magnetisation ($M$) for  $\lambda=0.3$ and $\lambda=0.7$ (b) 
Susceptibility ($\chi$) for  $\lambda=0.3$ and $\lambda=0.7$ (c) Specific heat
($C$) for  $\lambda=0.3$ and $\lambda=0.7$. In the right panel, for impurity
concentration $p=0.4$,
(d) Magnetisation ($M$) for  $\lambda=0.3$ and $\lambda=0.7$ (e) 
Susceptibility ($\chi$) for  $\lambda=0.3$ and $\lambda=0.7$ (f) Specific heat ($C$) for  $\lambda=0.3$ and $\lambda=0.7$.}

\label{all-T-lambda}
\end{center}
\end{figure}

Let us start by presenting the simulational outcomes , highlighting the findings obtained for bilinear exchange type of anisotropy ($\lambda$)
We have calculated the magnetisation ($M$), susceptibility ($\chi$) and the specific heat ($C$) and studied these quantities
as functions of the temperature ($T$). The temperature variations of all these quantities (for different values of bilinear exchange 
anisotropy $\lambda$ and impurity concentration $p$) are shown
in Fig-\ref{all-T-lambda}.

\vskip 0.3cm
%%%%%%%%%%%%%%%%%%%%%%%%%%%%%%%%%%%%%%%%%%%%%%%%%%%%%%%%%%%%%%%%%%%%%%%

\vskip 0.3cm

While cooling the system for a fixed set of values of the strength of bilinear exchange type of anisotropy ($\gamma$) and the
concentration of impurity ($p$), it is observed that the order parameter or the magnetisation ($M$) transits from zero to a nonzero value
continuously, indicating the continuous phase transition. The susceptibility ($\chi$) and specific heat ($C$) exhibit maxima at a particular temperature, signifying the occurrence of a phase transition. The pseudocritical temperature ($T_c^{*}$) is determined from  the temperature at which the susceptibility ($\chi$) attains its maximum value. Moreover, the pseudocritical temperature has been observed to change depending
on the value of $\lambda$ and $p$. For any fixed value of $\lambda$ the pseudocritical temperature ($T_c^*$) decreases as the
concentration of impurity ($p$) increases. It is pertinent to note here
that the decrease of the pseudocritical temperature ($T_c^{*}$) with the increase of the impurity concentration ($p$) has already been noticed
\cite{filho} in the three dimensional {\it isotropic} XY ferromagnet by Monte Carlo
simulation. This was confirmed later\cite{olivia-rev} and found the linear
dependence of pseudocritical temperature $T_c^{*}$ on the impurity concentration $p$. On the other hand, the pseudocritical temperature ($T_c^*$) increases with
$\lambda$ for the fixed value of the impurity concentration $p$. These are demonstrated in Fig-\ref{all-T-lambda}.

\vskip 0.2cm

%%%%%%%%%%%%%%%%%%%%%%%%%%%%%%%%%%%%%%%%%%%%%%%%%%%%%%%%%%%%%%%%%%%%%%%
\begin{figure}[h]
\begin{center}

\resizebox{10cm}{!}{\includegraphics[angle=0]{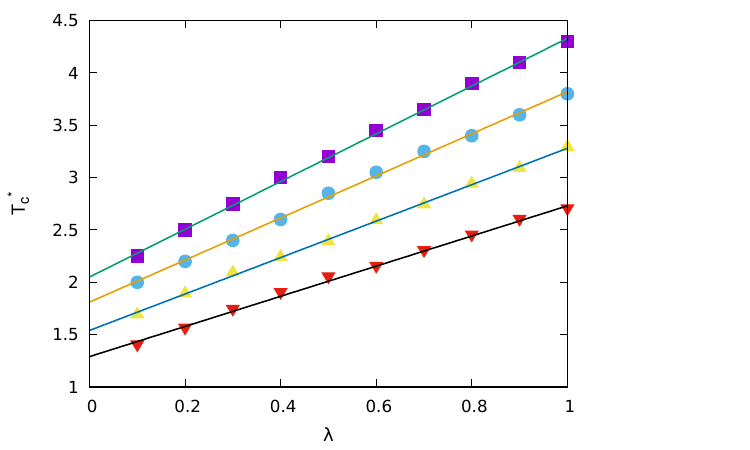}}

\caption{The pseudocritical temperature ($T_c^{*}$) is shown as a function of the strengths of the bilinear exchange type of anisotropy ($\lambda$) 
for different impurity concentrations
($p$) for. The solid lines represent corresponding linear best fit $y=a+bx$. $p=0.1$ (represented by Squares,
$a=2.05,b=2.28$), $p=0.2$ (represented by Bullets, $a=1.81, b=2.01$), $p=0.3$ (represented by Triangles, $a=1.54, b=1.74$)
and $p=0.4$ (represented by Inverted triangles, $a=1.29, b=1.44$).}
\label{Tc-lambda-diff-p}
\end{center}
\end{figure}

\vskip 0.2cm

\begin{figure}[h]
\begin{center}

\resizebox{10cm}{!}{\includegraphics[angle=0]{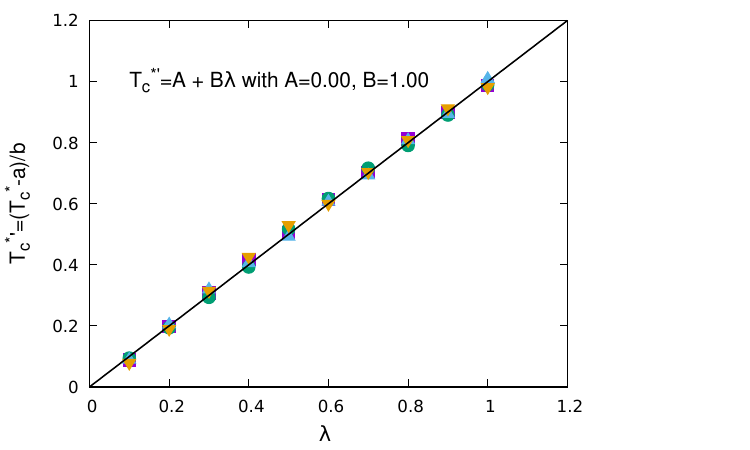}}

\caption{The scaled pseudocritical temperature ($T_c^{*'}=(T_c^{*}-a)/b$) is shown as a function of the anisotropy ($\lambda$).
The values of $a$ and $b$ are collected from Fig-\ref{Tc-lambda-diff-p}. The solid line is the fitted (linear best fit) function ($T_C^{*'}=A+B\lambda$) obtained from the best fit. Here, $A$=0.00 and $B$=1.00.}

\label{scaled-Tc-lambda-diff-p}
\end{center}
\end{figure}

We have tried to study the dependence of $T_c^*$ on $\lambda$ and $p$, systematically. Fig-\ref{Tc-lambda-diff-p} shows the
variation of $T_c^*$ as function of $\lambda$ for four different values of $p$. $T_c^*$ is found to be linear in $\lambda$.
The data show linear best fit $T_c^* = a + b \lambda$.
However, the slope ($b$) and the intercept ($a$) depends on $p$. The scaled pseudocritical temperature ($T_c^{*'}= (T_c^*-a)/b$)
has been plotted with $\lambda$. All data collapsed (Fig-\ref{scaled-Tc-lambda-diff-p}) on a straight line $T_c^{*'} = \lambda$.  

\vskip 0.2cm

%%%%%%%%%%%%%%%%%%%%%%%%%%%%%%%%%%%%%%%%%%%%%%%%%%%%%%%%%%%%%%%%%%%%%%%%%%

\vskip 0.2cm

\begin{figure}[h]
\begin{center}

\resizebox{10cm}{!}{\includegraphics[angle=0]{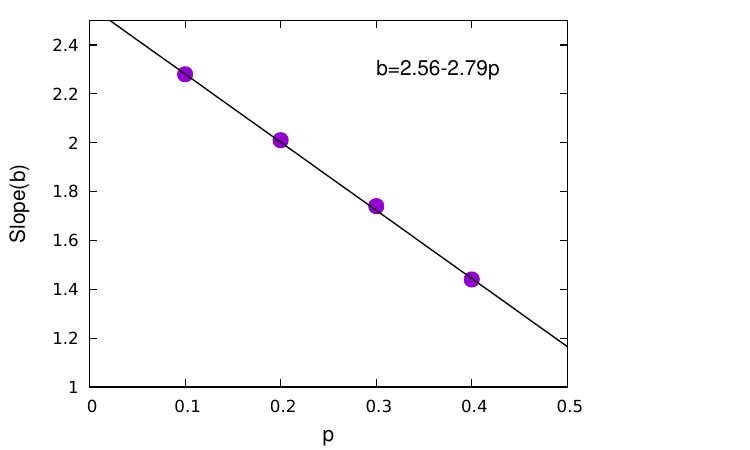}}

\caption{The slope of the fitted straight line (Fig-\ref{Tc-lambda-diff-p}) $b$ is plotted against different
impurity concentration $p$. The data are fitted to a straight line (solid line) $b=2.56-2.79p$.}

\label{slope-p-lambda-type}
\end{center}
\end{figure}

The slope ($b$) has been found to depend linearly ($b= A + B p$) on the impurity concentration ($p$). We have estimated $A=2.56$ and $B=-2.79$.
This is shown in Fig-\ref{slope-p-lambda-type}. This has not been studied before.

\vskip 0.2cm
%%%%%%%%%%%%%%%%%%%%%%%%%%%%%%%%%%%%%%%%%%%%%%%%%%%%%%%%%%%%%%%%%%%%%%%%%%

\vskip 0.2cm

\begin{figure}[h]
\begin{center}

\resizebox{10cm}{!}{\includegraphics[angle=0]{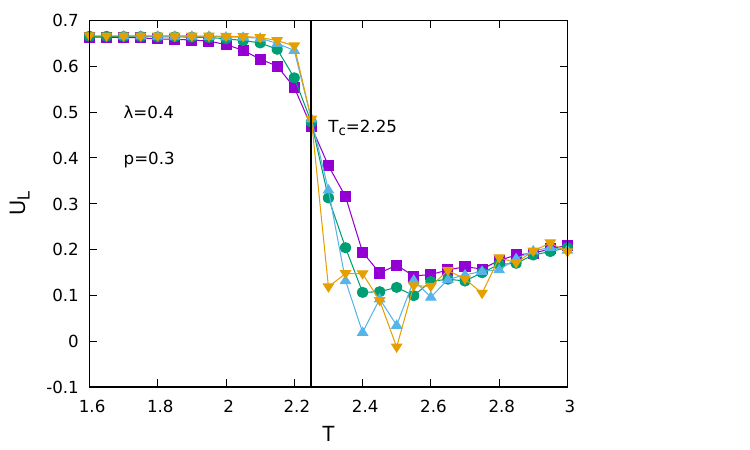}}

\caption{The fourth order Binder cumulant ($U_L$) is plotted against the temperature ($T$) for different system sizes ($L$) and fixed values
of impurity concentration ($p=0.3$) and strength of bilinear exchange anisotropy ($\lambda=0.4$). The intersection point indicates the critical
temperature ($T_c$), shown by the vertical black straight line. Here, $L=10$ (Square), $L=15$ (Bullet), $L=20$ (Triangle) and $L=25$
(Inverted triangle).}

\label{Binder-L-fixed-lambdap}
\end{center}
\end{figure}

We have also done a systematic finite size analysis for $L=10,15,20,25$ to substantiate the growth of critical correlation
(along with divergence of susceptibility)
at the phase transition temperature. The fourth order Binder cumulant has been studied as function of temperature with 
$L$ as parameter, for fixed $\lambda=0.4$ and $p=0.3$. This is shown in Fig-\ref{Binder-L-fixed-lambdap}. The intersection of all such curves determines the
true critical temperature $T_c$ (i.e., $T_c^*(L)$ for $L \to \infty$). Here, we have estimated $T_c=2.25$. 

\vskip 0.2cm
%%%%%%%%%%%%%%%%%%%%%%%%%%%%%%%%%%%%%%%%%%%%%%%%%%%%%%%%%%%%%%%%%%%%%%%%%%

\vskip 0.2cm

\begin{figure}[h]
\begin{center}

\resizebox{10cm}{!}{\includegraphics[angle=0]{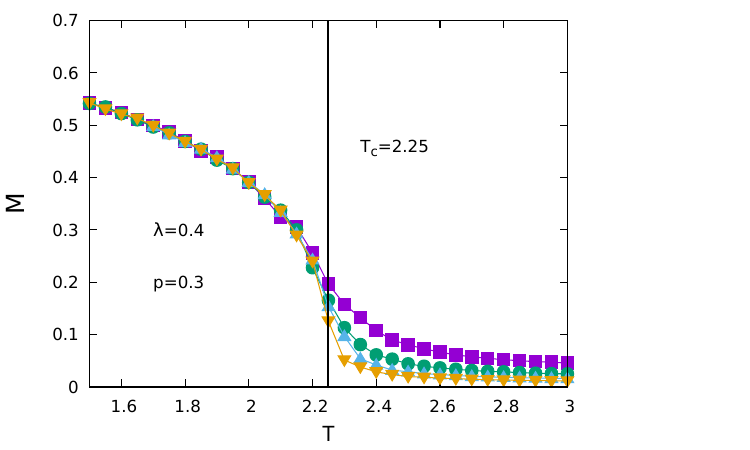}}

\caption{The magnetisation ($M$) is plotted against the temperature ($T$) for different system sizes ($L$) and fixed values
of impurity concentration ($p=0.3$) and strength of the bilinear exchange anisotropy ($\lambda=0.4$). The vertical black line indicates the critical temperature estimated from the intersection of Binder cumulant. The values of $M(T_c)$ are measured from the intersections of $M$ and the vertical line. Here, $L=10$ (Square), $L=15$ (Bullet), $L=20$ (Triangle) and $L=25$
(Inverted triangle).}

\label{magn-T-L-fixed-lambdap}
\end{center}
\end{figure}

We have also studied the temperature dependence of magnetisation ($M$) for different system sizes ($L=10,15,20,25$) for fixed
anisotropy $\lambda=0.4$ and  impurity concentration $p=0.3$. This is depicted in Fig-\ref{magn-T-L-fixed-lambdap}. The vertical line ($T_c=2.25$) intersects the curves at different points. 

\vskip 0.2cm
%%%%%%%%%%%%%%%%%%%%%%%%%%%%%%%%%%%%%%%%%%%%%%%%%%%%%%%%%%%%%%%%%%%%%%%%%%

\vskip 0.2cm

\begin{figure}[h]
\begin{center}

\resizebox{10cm}{!}{\includegraphics[angle=0]{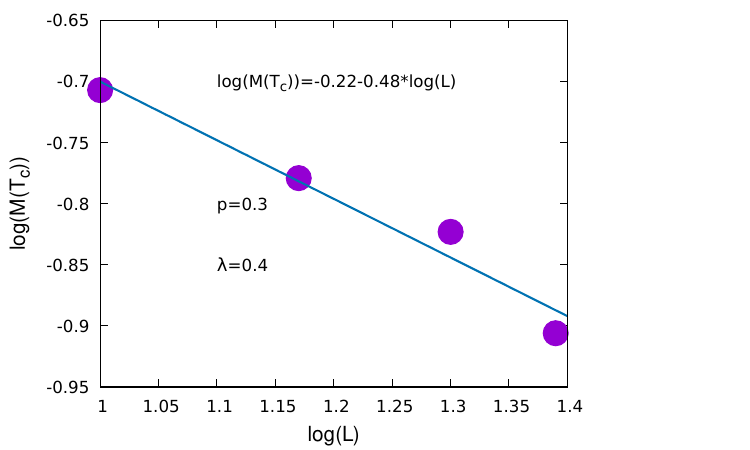}}

\caption{The logarithm of the values of critical magnetisation 
($log(M(T_c))$ is plotted for logarithm of different $L$. 
Here, $\lambda=0.4$ and $p=0.3$. The solid line is the best fit. The 
exponent ${{\beta} \over {\nu}} \sim 0.48\pm0.05$ for the scaling $M(T_c) \sim L^{-{{\beta} \over {\nu}}}$.}

\label{log(magn)-T-L-fixed-lambdap}
\end{center}
\end{figure}

From these intersections, 
the critical magnetisations ($M(T_c)$),
for various system sizes ($L=10,15,20,25$), are determined. Assuming the scaling law\cite{stanley}, $M(T_c) \sim L^{-{{\beta} \over {\nu}}}$,
the log$((M(T_c))$ is plotted against log($L$). This is shown in Fig-\ref{log(magn)-T-L-fixed-lambdap}. 

\vskip 0.2cm
%%%%%%%%%%%%%%%%%%%%%%%%%%%%%%%%%%%%%%%%%%%%%%%%%%%%%%%%%%%%%%%%%%%%%%%%%%

\vskip 0.2cm

This has been fitted with a straight line. The slope of the straight line results the scaling exponent
${{\beta} \over {\nu}}=0.48\pm0.05$.

\begin{figure}[h]
\begin{center}

\resizebox{10cm}{!}{\includegraphics[angle=0]{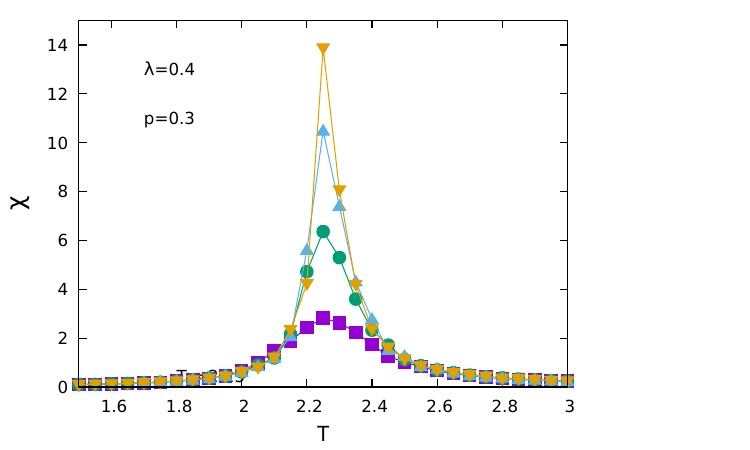}}

\caption{The  susceptibility 
($\chi_p)$ is plotted against the temperature ($T$) for different $L$. 
Here, $\lambda=0.4$ and $p=0.3$. The solid line is the best fit.Here, $L=10$ (Square), $L=15$ (Bullet), $L=20$ (Triangle) and $L=25$
(Inverted triangle).}

\label{chi-T-L-fixed-lambdap}
\end{center}
\end{figure}

The divergence of the susceptibility at the transition temperature is a crucial phenomenon in the equilibrium continuous phase
transitions. This can be studied by checking the increase of peak height of the susceptibility with increase of the
system size $L$. We have studied the temperature variation of $\chi$ for different system sizes ($L=10,15,20,25$) with fixed
anisotropy $\lambda=0.4$ and impurity concentration $p=0.3$. Fig-\ref{chi-T-L-fixed-lambdap} shows that peak height of the
susceptibility ($\chi_p$) increases as the system size $L$ increases. 

\vskip 0.2cm
%%%%%%%%%%%%%%%%%%%%%%%%%%%%%%%%%%%%%%%%%%%%%%%%%%%%%%%%%%%%%%%%%%%%%%%%%%

\vskip 0.2cm

\begin{figure}[h]
\begin{center}

\resizebox{10cm}{!}{\includegraphics[angle=0]{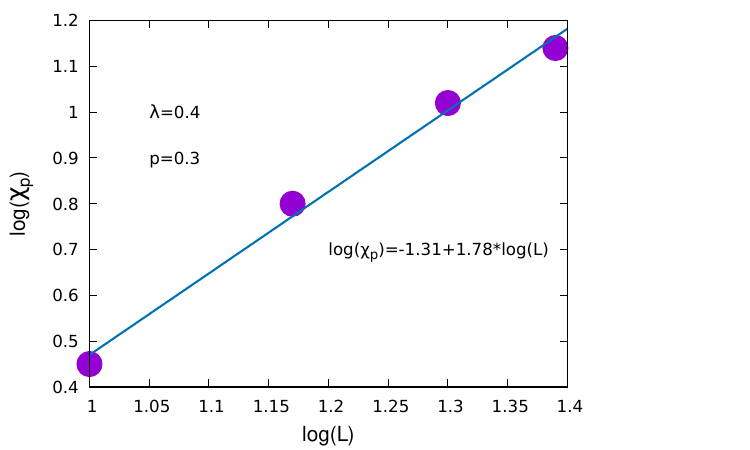}}

\caption{The logarithm of the values of maximum susceptibility 
($log(\chi_p))$ is plotted for logarithm of different $L$. 
Here, $\lambda=0.4$ and $p=0.3$. The solid line is the best fit. The 
exponent ${{\gamma} \over {\nu}} \sim 1.78\pm0.05$ for the scaling $\chi_p \sim L^{{\gamma} \over {\nu}}$.}

\label{log(chipeak)-T-L-fixed-lambdap}
\end{center}
\end{figure}

This indicates that the susceptibility would eventually
diverge for $L \to \infty$. Here also, assuming the scaling law\cite{stanley}  $\chi_p \sim L^{{{\gamma} \over {\nu}}}$, we have plotted
log($\chi_p$) with log($L$) for fixed $\lambda=0.4$ and $p=0.3$. This is shown in Fig-\ref{log(chipeak)-T-L-fixed-lambdap}.

\vskip 0.2cm
%%%%%%%%%%%%%%%%%%%%%%%%%%%%%%%%%%%%%%%%%%%%%%%%%%%%%%%%%%%%%%%%%%%%%%%%%%

\vskip 0.2cm

This has been fitted with a straight line.
The slope of the straight line estimates the scaling exponent ${{\gamma} \over {\nu}}=1.78\pm0.05$. These finite size analysis
confirms the transition as the true thermodynamic phase transition.

\newpage

\noindent {\bf {\it {(b). Single site Anisotropy ($D$)}}}

\vskip 0.3cm

\begin{figure}[h]
\begin{center}

\resizebox{10cm}{!}{\includegraphics[angle=0]{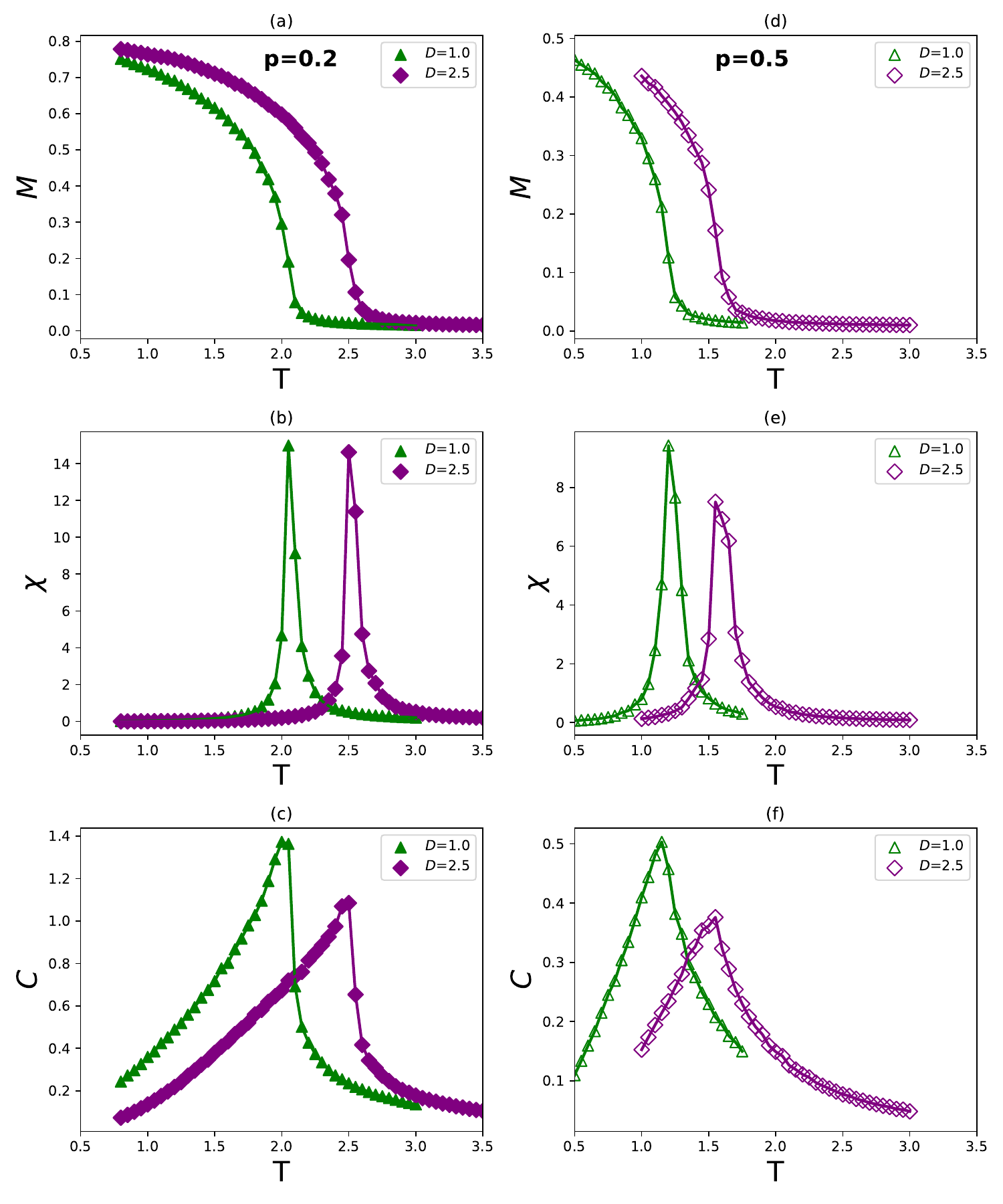}}

\caption{The temperature dependences of all thermodynamic quantities for 
single site anisotropy.
In the left panel, for impurity concentration $p=0.2$, 
(a) Magnetisation ($M$) for  $D=1.0$ and $D=2.5$ (b) 
Susceptibility ($\chi$) for  $D=1.0$ and $D=2.5$ (c) Specific heat
($C$) for  $D=1.0$ and $D=2.5$. In the right panel, for impurity
concentration $p=0.5$,
(d) Magnetisation ($M$) for  $D=1.0$ and $D=2.5$ (e) 
Susceptibility ($\chi$) for  $D=1.0$ and $D=2.5$ (f) Specific heat
($C$) for  $D=1.0$ and $D=2.5$.}
\label{all-T-D}
\end{center}
\end{figure}

In this subsection, we report the simulational results of our study for the single site anisotropy ($D$).
Here also, we have studied the temperature variation of the magnetisation ($M$), susceptibility ($\chi$) and the specific heat ($C$). All these
quantities are shown in Fig-\ref{all-T-D} as functions of temperature for
different values of single site anisotropy $D$ and impurity concentration
$p$.

\vskip 0.2cm
%%%%%%%%%%%%%%%  FIGURES For D %%%%%%%%%%%%%%%%%%%%%%%%%%%%%%%%%%%%
%%%%%%%%%%%%%%%%%Figure-10%%%%%%%%%%%%%%%%%%%%%%%%%%%%%%%%%%%%%%%%%%

\vskip 0.2cm

\begin{figure}[h]
\begin{center}

\resizebox{14cm}{!}{\includegraphics[angle=0]{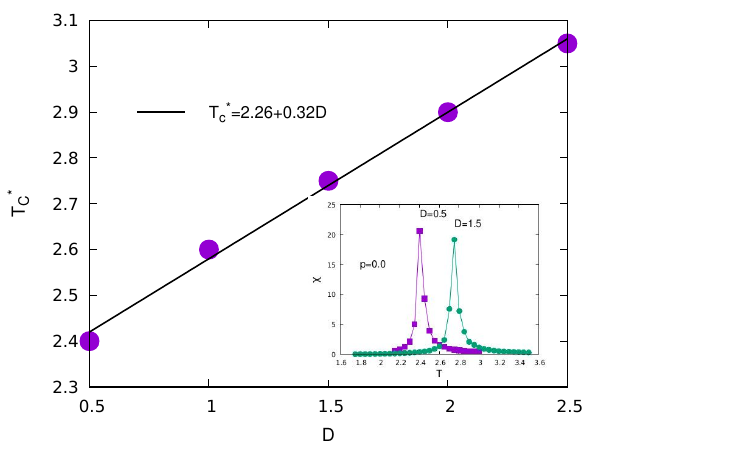}}

\caption{The pseudocritical temperature ($T_c^{*}$) is plotted against
the strength of single-site anisotropy ($D$) for pure ($p=0.0$) system.
The inset shows the temperature ($T$) dependences of the suscptibility
($\chi$) for two different values of $D$.}
\label{Tc-D-p0-full}
\end{center}
\end{figure}

The results of pseudocritical temperature ($T_C^*$) as function of single site anisotropy ($D$) for pure ($p=0$) system are shown in Fig.\ref{Tc-D-p0-full}. 

\vskip 0.2cm
%%%%%%%%%%%%%%%  Figure-11%%%%%%%%%%%%%%%%%%%%%%%%%%%%%%%%%%%%%%%%%%

\vskip 0.2cm

This is
first Monte Carlo results. This was initially done by a finite temperature quantum field theoretic calculation\cite{ma}. The Monte Carlo results 
show a linear variation $T_c^*=A+B D$ with $A=2.26$ and $B=0.32$ which
is significantly different from that obtained in quatum calculations\cite{ma}.  It may be noted here that the pseudocritical temperature ($T_c^*$) increases linearly\cite{olivia1} with the strength of anisotropy ($\lambda$) in the case of bilinear exchange anisotropy. 

While cooling the system for fixed set of values (Fig-\ref{all-T-D}) of the strength of the single site  anisotropy ($D$) and the
concentration of impurity ($p$), it is observed that the order parameter or the magnetisation ($M$) shows a continuous variation,  from a zero to a non-zero value, indicative of the continuous phase transition. The susceptibility ($\chi$) and the specific heat ($C$) get peaked
at any temperature indicating the phase transition. The pseudocritical temperature ($T_c^*$) is calculated from the temperature
which maximises the susceptibility ($\chi$). Moreover, the pseudocritical temperature ($T_c^{*}$) depends on the values of $D$ and $p$. For any fixed value of $D$ the pseudocritical temperature ($T_c^*$) decreases as the
concentration of impurity ($p$) increases. Conversely, the pseudocritical temperature ($T_c^*$) has been found to increase 
with the increase of $D$ for a fixed value of the impurity concentration $p$.

\vskip 0.2cm
%%%%%%%%%%%%%%%%%%%%%%%%%Figure-12%%%%%%%%%%%%%%%%%%%%%%%%%%%%
\begin{figure}[h]
\begin{center}

\resizebox{10cm}{!}{\includegraphics[angle=0]{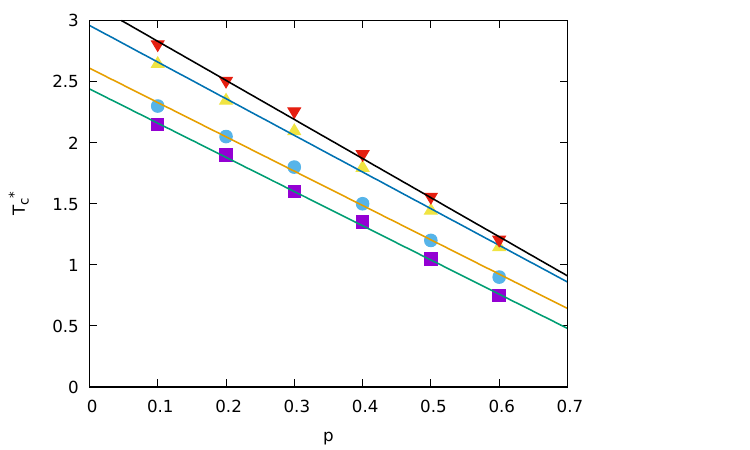}}

\caption{The pseudocritical temperature ($T_c^{*}$) is shown as a function of impurity concentration
($p$) for different strengths of anisotropy ($D$). The solid lines represent corresponding linear best fits $y=a-bx$. $D=0.5$ (represented by Squares,
$a=2.44,b=2.80$), $D=1.0$ (represented by Bullets, $a=2.61, b=2.81$), $D=2.0$ (represented by Triangles, $a=2.96, b=3.00$)
and $D=2.5$ (represented by Inverted triangles, $a=3.15, b=3.20$)}
\label{Tc-p-diff-D}
\end{center}
\end{figure}

\vskip 0.2cm

\begin{figure}[h]
\begin{center}

\resizebox{10cm}{!}{\includegraphics[angle=0]{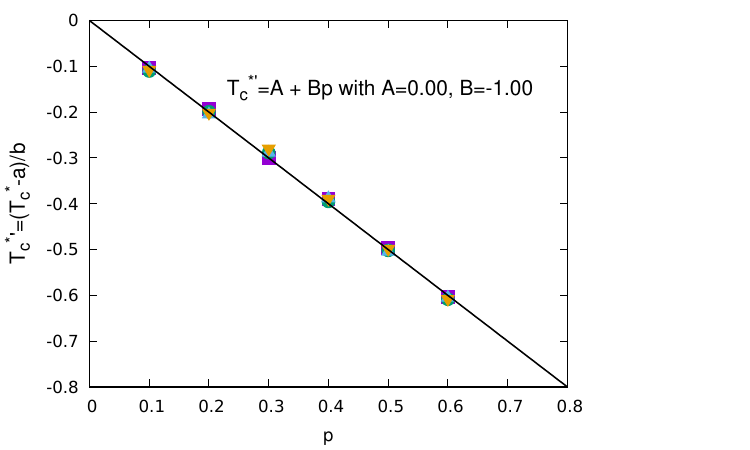}}

\caption{The scaled pseudocritical temperature ($T_c^{*'}=(T_c^{*}-a)/b$) is shown as a function of the concentration of impurity ($p$).
The values of $a$ and $b$ are collected from Fig-\ref{Tc-p-diff-D}. The solid line is the fitted (linear best fit) function ($T_C^{*'}=A+B p$) obtained from the best fit. Here, $A$=0.00 and $B$=-1.00.}

\label{scaled-Tc-p-diff-D}
\end{center}
\end{figure}

We have tried to study the dependence of $T_c^*$ on $D$ and $p$, systematically. Fig-\ref{Tc-p-diff-D} shows the
variation of $T_c^*$ as function of $p$ for four different values of anisotropy $D$. Here also, $T_c^*$ is found to be linear in $p$.
The data show linear best fit $T_c^* = a - bp$. Unlike the case of bilinear exchange kind of anisotropy, we failed to
make any conclusive remark regarding the dependence of the slope $b$ on the anisotropy $D$.
However, the slope ($b$) and the intercept ($a$) depends on $D$. The scaled speudocritical temperature ($T_c^{*'}= (T_c^*-a)/b$)
has been plotted with $p$. All data collapsed on a single straight line $T_c^{*'} = p$ and shown in Fig-\ref{scaled-Tc-p-diff-D}. 

\newpage

%%%%%%%%%%%%%%%%%%Figure-13%%%%%%%%%%%%%%%%%%%%%%%%%%%%%%%%%%%%%%%%%

\vskip 0.2cm

\begin{figure}[h]
\begin{center}

\resizebox{10cm}{!}{\includegraphics[angle=0]{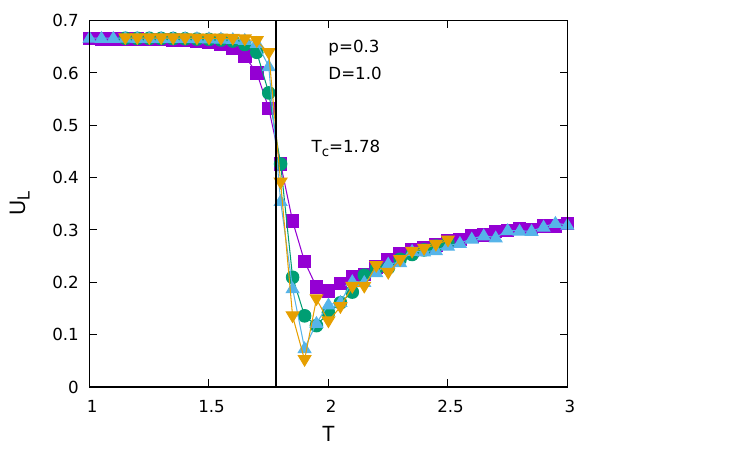}}

\caption{The fourth order Binder cumulant ($U_L$) is plotted against the temperature ($T$) for different system sizes ($L$) and fixed values
of impurity concentration ($p=0.3$) and strength of single site anisotropy ($D=1.0$). The intersection point indicates the critical
temperature ($T_c$), shown by the vertical black straight line. Here, $L=10$ (Square), $L=15$ (Bullet), $L=20$ (Triangle) and $L=25$
(Inverted triangle).}

\label{Binder-L-fixed-Dp}
\end{center}
\end{figure}

We have also done a systematic finite size analysis for $L=10,15,20,25$ to check the growth of critical correlation
(along with divergence of susceptibility)
at the phase transition temperature. The fourth order Binder cumulant has been studied as function of temperature with 
$L$ as parameter, for fixed $D=1.0$ and $p=0.3$. This is shown in Fig-\ref{Binder-L-fixed-Dp}. 

\vskip 0.2cm
%%%%%%%%%%%%%%%Figure-14%%%%%%%%%%%%%%%%%%%%%%%%%%%%%%%%%%%%%%%%%%%%

\vskip 0.2cm

The intersection of all such curves determines the
true critical temperature $T_c$ (i.e., $T_c^*(L)$ for $L \to \infty$). Here, we have estimated $T_c=1.78$. 

\newpage

%%%%%%%%%%%%%%%%%Figure-15%%%%%%%%%%%%%%%%%%%%%%%%%%%%%%%%%%%%%%%%%%%%%
\begin{figure}[h]
\begin{center}

\resizebox{10cm}{!}{\includegraphics[angle=0]{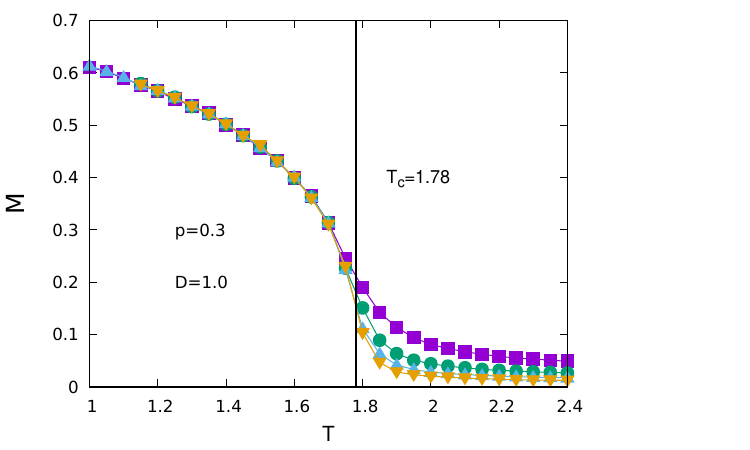}}

\caption{The magnetisation ($M$) is plotted against the temperature ($T$) for different system sizes ($L$) and fixed values
of impurity concentration ($p=0.3$) and strength of single site anisotropy ($D=1.0$). The vertical black line indicates the critical temperature estimated from the intersection of Binder cumulant. The values of $M(T_c)$ are measured from the intersections of $M$ and the vertical line. Here, $L=10$ (Square), $L=15$ (Bullet), $L=20$ (Triangle) and $L=25$
(Inverted triangle).}

\label{magn-T-L-fixed-Dp}
\end{center}
\end{figure}

\vskip 0.2cm

\begin{figure}[h]
\begin{center}

\resizebox{10cm}{!}{\includegraphics[angle=0]{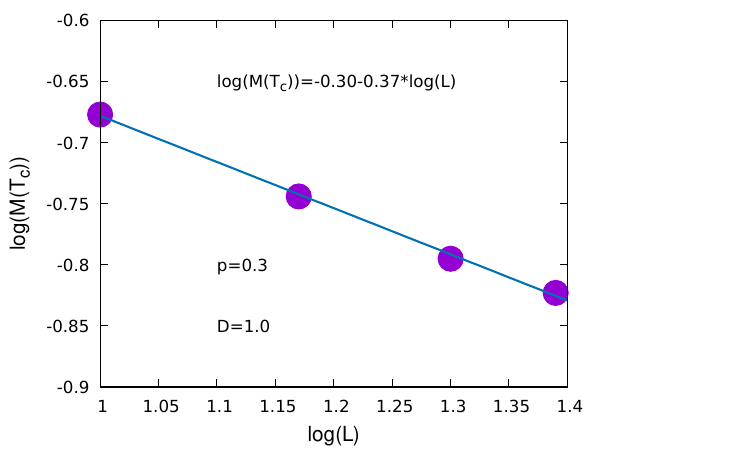}}

\caption{The logarithm of the values of critical magnetisation 
($log(M(T_c))$ is plotted for logarithm of different $L$. 
Here, $D=1.0$ and $p=0.3$. The solid line is the best fit. The 
exponent ${{\beta} \over {\nu}} \sim 0.37\pm0.05$ for the scaling $M(T_c) \sim L^{-{{\beta} \over {\nu}}}$.}

\label{log(magn)-T-L-fixed-Dp}
\end{center}
\end{figure}

We have also studied the temperature dependence of magnetisation ($M$) for different system sizes ($L=10,15,20,25$) for fixed
anisotropy $D=1.0$ and  impurity concentration $p=0.3$. This has been shown in Fig-\ref{magn-T-L-fixed-Dp}. The vertical line ($T_c=1.78$) intersects the curves at different points. From these intersections, 
the critical magnetisations ($M(T_c)$),
for various system sizes ($L=10,15,20,25$), are determined. Assuming the scaling law\cite{stanley}, $M(T_c) \sim L^{-{{\beta} \over {\nu}}}$,
the log$((M(T_c))$ is plotted against log($L$). This is shown in Fig-\ref{log(magn)-T-L-fixed-Dp}. 

\newpage

%%%%%%%%%%%%%%%%%%Figure-16%%%%%%%%%%%%%%%%%%%%%%%%%%%%%%%%%%%%%%%%%%%%%

\newpage

This has been fitted with a straight line. The slope of the straight line results the scaling exponent
${{\beta} \over {\nu}}=0.37\pm0.04$.

\begin{figure}[h]
\begin{center}

\resizebox{10cm}{!}{\includegraphics[angle=0]{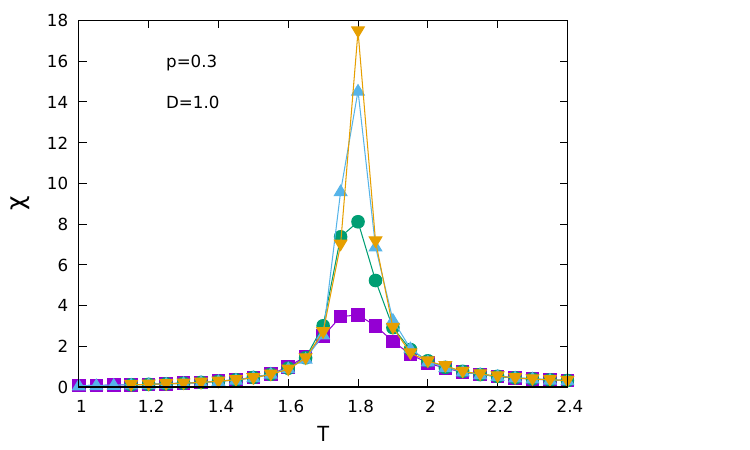}}

\caption{The  susceptibility 
($\chi_p)$ is plotted against the temperature ($T$) for different $L$. 
Here, $D=1.0$ and $p=0.3$. The solid line is the best fit.Here, $L=10$ (Square), $L=15$ (Bullet), $L=20$ (Triangle) and $L=25$
(Inverted triangle).}

\label{chi-T-L-fixed-Dp}
\end{center}
\end{figure}

The divergence of the susceptibility at the transition temperature is a crucial phenomenon in the equilibrium continuous phase
transitions. This can be examined by checking the increase of peak height of the susceptibility with increase of the
system size $L$. We have studied the temperature variation of $\chi$ for different system sizes ($L=10,15,20,25$) with fixed
anisotropy $D=1.0$ and impurity concentration $p=0.3$. Fig-\ref{chi-T-L-fixed-Dp} shows that peak height of the
susceptibility ($\chi_p$) increases as the system size $L$ increases. 

\newpage

%%%%%%%%%%%%%Figure-17%%%%%%%%%%%%%%%%%%%%%%%%%%%%%%%%%%%%%%%%%%%%%%

\begin{figure}[h]
\begin{center}

\resizebox{10cm}{!}{\includegraphics[angle=0]{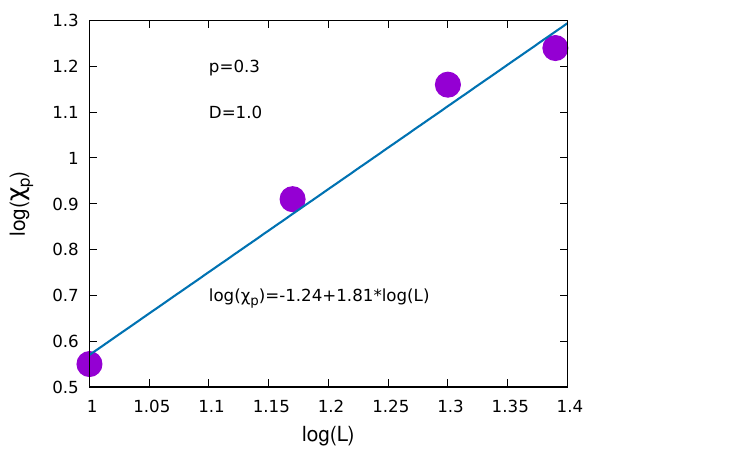}}

\caption{The logarithm of the values of maximum susceptibility 
($log(\chi_p))$ is plotted for logarithm of different $L$. 
Here, $D=1.0$ and $p=0.3$. The solid line is the best fit. The 
exponent ${{\gamma} \over {\nu}} \sim 0.1.81\pm0.05$ for the scaling $\chi_p \sim L^{{\gamma} \over {\nu}}$.}

\label{log(chipeak)-T-L-fixed-Dp}
\end{center}
\end{figure}

This indicates that the susceptibility would eventually
diverge for $L \to \infty$. Here also, assuming the scaling law\cite{stanley}  $\chi_p \sim L^{{{\gamma} \over {\nu}}}$, we have plotted
log($\chi_p$) with log($L$) for fixed $D=1.0$ and $p=0.3$. This is shown in Fig-\ref{log(chipeak)-T-L-fixed-Dp}.

%\newpage

%%%%%%%%%%%%%%%Figure-18%%%%%%%%%%%%%%%%%%%%%%%%%%%%%%%%%%%%%%%%%%%%%%
 
%\newpage

This has been fitted with a straight line.
The slope of the straight line estimates the scaling exponent ${{\gamma} \over {\nu}}=1.81\pm0.05$. These finite size analysis
validates the transition as the true thermodynamic phase transition.
 
\vskip 1cm

\newpage

\noindent {\bf {\Large V. Summary}}

\vskip 0.2cm

\noindent The role of nonmagnetic impurity on the critical behaviours of the anisotropic XY ferromagnet has not been
studied before. We have tried to explore such role of nonmagnetic impurity on the anisotropic XY ferromagnet
through a systematic investigation via Monte Carlo simulation in three dimensions with Metropolis algorithm.
We have considered here two different kinds of anisotropy, namely, the bilinear exchange type and the single site
anisotropy. 

Our major findings are, in the case of bilinear exchange anisotropy, the pseudocritical temperature of the
ferro-para phase transition increases linearly with the strength of anisotropy. However, the slope ($b$) of such
linear increase, decreases linearly as the concentration ($p$) of the nonmagnetic impurity increases. The critical exponents for the assumed scaling laws $M(T_c) \sim L^{-{{\beta} \over {\nu}}}$ and $\chi_p \sim L^{{{\gamma} \over {\nu}}}$ are estimated through the finite size analysis. We have estimated, 
${{\beta} \over {\nu}}$ equals to  $0.48\pm0.05$ and ${{\gamma} \over {\nu}}$ equals to $1.78\pm0.05$. It may be noted here that
the exponent ${{\beta} \over {\nu}}$ is very close to ${{1} \over {2}}$. 

On the other hand, for single site anisotropy, the pseudocritical temperature has been found to decrease linearly with
the impurity concentration. The slope of this linear fall depends on the single site anisotropy ($D$). However, we failed
to extract any systematic dependence of it. In the case of single site anisotropy, the critical exponents for the assumed scaling laws $M(T_c) \sim L^{-{{\beta} \over {\nu}}}$ and $\chi_p \sim L^{{{\gamma} \over {\nu}}}$ are estimated through the systematic and usual finite size analysis. We have estimated, 
${{\beta} \over {\nu}}$ equals to  $0.37\pm0.04$ and ${{\gamma} \over {\nu}}$ equals to $1.81\pm0.05$. It may be noted here that
the exponent ${{\beta} \over {\nu}}$ is very close to ${{1} \over {3}}$. 

The scaling exponents ${{\gamma} \over {\nu}}$, for both bilinear exchange anisotropy and single site anisotropy, provides
almost same value. But the values of other exponent ${{\beta} \over {\nu}}$, differ significantly for bilinear exchange and
single site anisotropy. This prompted us to conclude that these two kinds of anisotropy belong to two different universality
classes\cite{stanley}.

As far as the knowledge of these authors is concerned, this is the first
study of the effects of randomly quenched disorder (of nonmagnetic impurity) in the phase transition of three dimensional {\it anisotropic} planar ferromagnet. The main conclusion of this study is that the two types of anisotropy (bilinear exchange type or single site type) belong to two different universality classes. The universality class generally depends on
the dimensionality of the system and the symmetry of the order parameter
in the case of equilibrium phase transitions. But in the present study,
as the results show, that the types of anisotropy may lead to different universality classes. Recently, the breaking of the same universality class for the site percolation and bond percolation has been reported in the
statistical physics of networks\cite{network}.

\vskip 1cm

\noindent {\bf {\Large Acknowledgements:}}  OM acknowledges MANF,UGC, Govt. of India for financial support. MA acknowledges FRPDF
of Presidency University, Kolkata.

\vskip 1cm

\noindent {\bf Data availability statement:} Data will be available on request to Olivia Mallick.

\vskip 0.5cm

\noindent {\bf Conflict of interest statement:} We declare that this manuscript is free from any conflict of interest. The authors have no financial or proprietary interests in any material discussed in this article.

\vskip 0.5cm

\noindent {\bf Funding statement:} No funding was received particularly to support this work.

%\newpage

\end{document}